\documentclass{article}

\input{setup.sty}

\begin{document}
\onehalfspacing

\begin{center}

~
\vskip5mm

{\LARGE  {
Replica manifolds, pole skipping, and the butterfly effect
}}

\vskip7mm 

Wan Zhen Chua, Thomas Hartman and Wayne W. Weng

\vskip5mm

{\it Department of Physics, Cornell University, Ithaca, New York, USA
}

\vskip5mm

\end{center}

\vspace{2mm}

\begin{abstract}
\noindent The black hole butterfly effect is a signal of quantum chaos in holographic theories that can be probed in different ways, including out-of-time-order correlators (OTOCs), pole skipping (PS), and entanglement wedge (EW) reconstruction. Each of these three phenomena can be used to define a butterfly velocity that measures the speed at which chaos spreads. In a general quantum system the three velocities $\vbo$, $\vbp$, and $\vbe$ can be different, but it is known from explicit calculations that they are all equal in certain holographic theories dual to Einstein gravity plus higher-curvature corrections. A conceptual explanation for this apparent coincidence is lacking. We show that it follows from a deeper relationship: The pole-skipping mode, added to the black hole background, can be reinterpreted as the gravitational replica manifold for the late-time entanglement wedge, and its imaginary part is the shockwave that computes the OTOC. Thus pole skipping is directly related to entanglement dynamics in holographic theories, and the origin of the pole-skipping mode is an extremal surface on the horizon. This explains the coincidence $\vbo = \vbp = \vbe$ in known cases, and extends it to general theories of gravity with a pole-skipping mode having the usual behavior.


 \end{abstract}

\pagebreak
\pagestyle{plain}

\setcounter{tocdepth}{2}
{}
\vfill
\tableofcontents


\date{}

\section{Introduction}

Black holes are strongly chaotic quantum systems. This manifests in many ways, including fast scrambling, energy level repulsion, and the butterfly effect, which refers to the exponential sensitivity of a quantum state to small perturbations  \cite{Hayden:2007cs,Sekino:2008he,Shenker:2013pqa,Roberts:2014isa,Kitaev:2014talk,Cotler:2016fpe}. In the butterfly effect, chaos grows in time at a rate determined by the quantum Lyapunov exponent, $\lambda_L$, and spreads in space at a speed known as the butterfly velocity, $v_B$.

There is no single definition of quantum chaos, and in general, different observables can have different Lyapunov exponents and butterfly velocities. Yet in a broad class of holographic theories, there is a curious coincidence \cite{Mezei:2016wfz, Dong:2022ucb}: The butterfly velocity defined by out-of-time-order thermal four-point correlation functions (OTOCs) is equal to the butterfly velocity defined by entanglement wedge (EW) reconstruction, $\vbo = \vbe$. This coincidence was first observed in theories dual to Einstein gravity plus curvature-squared corrections \cite{Mezei:2016wfz}, and later extended to all bulk theories with $f(\mbox{Riemann})$ Lagrangian \cite{Dong:2022ucb}. The calculations of $\vbo$ and $\vbe$ are \textit{a priori} completely different, and it was only after a long explicit calculation that they turned out to agree. (See also \cite{Baishya:2024sym}.)  All of these theories are maximally chaotic in the sense that they saturate the MSS chaos bound \cite{Maldacena:2015waa}.

In this paper, we will demonstrate a direct relationship between gravitational shockwaves and replica manifolds, in a limit where the replica manifold is defined by branching around a cross-section of a black hole horizon. Shockwaves are used to calculate OTOCs, and replica manifolds are used to find entanglement wedges. An immediate consequence of our construction is that $\vbo = \vbe$ in a general class of higher-derivative theories with a finite number of derivatives. At no point will we need to calculate either side explicitly, so this provides a conceptual explanation for the coincidence noted in \cite{Mezei:2016wfz,Dong:2022ucb}. It also gives a new perspective on why pole-skipping occurs in the first place, and naturally connects to the identification of the pole-skipping mode as a soft mode on the horizon \cite{Knysh:2024asf}.

To derive our result, we rely on yet another way to study the butterfly effect, using the phenomenon of pole skipping (PS) \cite{Grozdanov:2017ajz,Blake:2017ris,Blake:2018leo,Blake:2019otz,Grozdanov:2018kkt,Grozdanov_2019,Natsuume:2019sfp,Natsuume_2020,Natsuume:2019vcv,Ahn:2019rnq,wu:2019esr,Choi:2020tdj,Ramirez:2020qer,Blake_2021,Grozdanov_2021,Blake:2021hjj,Yadav:2023hyg,Baishya:2023mgz,Wang:2022mcq,Ning:2023ggs,Loganayagam_2023,Grozdanov_2023,Grozdanov:2023e,Ahn_2024,Natsuume:2023nonbh,Natsuume:2023miss}. Pole skipping is a universal feature of thermal two-point functions that occurs in theories where chaos is controlled by a hydrodynamic mode \cite{Grozdanov:2017ajz,Blake:2017ris,Blake_2021,Knysh:2024asf}. This defines a third butterfly velocity, $\vbp$, which could in principle differ from the other two. In holographic theories dual to higher-curvature gravity, possibly coupled to certain matter fields, it is known that $\vbo = \vbp$, because the gravitational shockwave and the pole-skipping mode satisfy the same equation of motion in the transverse directions \cite{Grozdanov:2017ajz,Blake:2018leo,Grozdanov:2018kkt,Ahn:2019rnq,Blake:2021hjj,Dong:2022ucb,Wang:2022mcq,Ning:2023ggs,Baishya:2024sym,wu:2019esr}.

Our main new contribution is to observe that the pole-skipping mode is identical to the displacement mode for the entanglement wedge of a large boundary region at late times, 
under a displacement along the butterfly cone. Equivalently, the pole-skipping mode, added to the black hole background, is a gravitational replica manifold branched around a cross-section of the black hole horizon that is shifted slightly away from the bifurcation surface.

Gravitational replica manifolds \cite{Lewkowycz:2013nqa} are solutions to the equations of motion that obey prescribed boundary conditions at infinity corresponding to $n$ copies of the dual quantum system, which appear in the derivation of the Ryu-Takayanagi formula \cite{Ryu:2006bv,Ryu:2006ef} and its generalizations \cite{Hubeny:2007xt}. The relation between the replica manifold and the pole-skipping mode proves that pole skipping and entanglement wedge reconstruction have the same butterfly velocity, $\vbp = \vbe$, and combined with the results of \cite{Wang:2022mcq, Ning:2023ggs} showing $\vbo = \vbp$, we have the triple equality
\begin{align}\label{introtriple}
\vbo = \vbp = \vbe. 
\end{align}
The result applies to any bulk theory of gravity with a finite number of higher-derivative corrections and minimally-coupled matter fields of low spin, under the assumption that the theory admits a pole-skipping mode of the same form seen in many examples. (The precise assumption is spelled out in section \ref{sec:psshock}.)

Underlying this equation is the more fundamental fact that the three seemingly distinct bulk spacetimes --- the shockwave, the pole-skipping mode, and the replica manifold --- are actually related in a simple way. There is a special extremal surface on the black hole horizon described in section \ref{sec:ps_equals_displacement} corresponding to the limiting entanglement surface for large regions at late times. Let $\delta ds_n^2$ be the correction to the black hole metric obtained by constructing a replica manifold branched around this extremal surface and expanding in the replica limit $n \to 1$, and denote the pole-skipping mode by $\delta ds^2_{\rm PS}$. These are two solutions to the linearized gravitational equations of motion. We will show that they are in fact identical, 
\begin{align}\label{eq:intro_modes_eq}
\delta ds_n^2 = (n-1) \delta ds^2_{\rm PS},
\end{align}
to leading order in $(n-1)$ as well as the displacement from the bifurcation surface. Thus in holographic theories, pole skipping is a direct consequence of the existence of the late-time replica manifold. We will demonstrate \eqref{eq:intro_modes_eq} by starting with the pole-skipping mode and providing an explicit change of coordinates to show that it is also a valid replica manifold. 

The extremal surface used in this construction is one that lies entirely on the event horizon, and is shifted infinitesimally away from the bifurcation surface. This infinitesimal null shift corresponds to a displacement of the boundary region along the butterfly cone. Thus $\delta ds^2_{n}$ in \eqref{eq:intro_modes_eq} is (the late-time limiting value of) the holographic dual of the displacement operator for a shift along the butterfly cone. In the case of three-dimensional gravity, where the planar black hole can be mapped to the vacuum state by a conformal transformation and $v_B = 1$, the result \eqref{eq:intro_modes_eq} is a special case of the holographic dual of the vacuum R\'enyi displacement operator derived in \cite{Balakrishnan:2016ttg}, as we show in section \ref{sec:BTZ}.

In radial gauge, one component of the replica metric is singular at the black hole horizon and can be regulated by an $i\varepsilon$ prescription; this causes it to pick up an imaginary part which is in fact a gravitational shockwave on the horizon, leading to the second key equation
\begin{align}\label{introshock}
\mbox{Im\ } \delta ds^2_{n} = \pi (n-1) \delta ds^2_{\rm shock},
\end{align}
where $\delta ds^2_{\rm shock}$ is the perturbation corresponding to the shockwave. This implies an operator relation that underlies the coincidence of the butterfly velocities. We will not explore the operator statement in detail in this work, but it is presumably related to the hydrodynamic theory of chaos developed in \cite{Blake:2017ris,Blake:2018leo,Grozdanov:2017ajz}. Our results also tie pole-skipping directly to the dynamics of entanglement, and lend support to the suggestion in \cite{Blake:2017ris,Blake:2018leo,Grozdanov:2017ajz} that pole-skipping is a universal smoking gun for the hydrodynamic origin of chaos. 

The derivation of these results is simple compared to the explicit calculations of $\vbo$ and $\vbe$ in higher-derivative gravity in \cite{Mezei:2016wfz, Dong:2022ucb}. We just need to take the late-time limit of the replica manifold, so that it approaches the horizon, and exhibit a coordinate change to the usual pole-skipping mode. Once we have set up the necessary background this can be done quite easily.

The outline of the paper is as follows. In section \ref{sec:butterfly} we review the three ways to define and calculate the butterfly velocity in holographic theories: shockwaves, pole skipping, and entanglement wedge reconstruction. In section \ref{sec:psshock} we review the results of \cite{Wang:2022mcq,Ning:2023ggs} establishing $\vbp=\vbo$ in higher derivative gravity and give a variant of their argument relating the shockwave to the imaginary part of the pole-skipping perturbation. We also discuss the pole-skipping mode away from the usual near-horizon limit. In section \ref{sec:ps_equals_displacement}, we construct the late-time replica manifold corresponding to the entanglement wedge of an infalling particle. This leads to the derivation of the main results. In section \ref{sec:BTZ}, we work out the example of the planar BTZ black hole, which is a special case where our results can be mapped to those of \cite{Balakrishnan:2016ttg}.

\section{Review of the butterfly effect in holographic theories}\label{sec:butterfly}

In this section, we review the butterfly effect as it appears in OTOCs, pole skipping and entanglement wedge reconstruction in holographic theories. We will state most of the results without derivation and refer to the original papers for details.

Throughout the paper, we assume the bulk theory is Einstein gravity plus a finite set of higher-derivative corrections (the `gravity limit'), which are treated perturbatively to all orders in the higher-derivative couplings, and possibly with minimally-coupled matter fields of spin less than two. These theories have maximal Lyapunov exponent, $\lambda_L = \frac{2\pi}{\beta}$, with $\beta$ the inverse temperature \cite{Shenker:2014cwa,Kitaev:2014talk,Maldacena:2015waa}. In all cases, we will assume the background geometry is an eternal black hole in AdS$_{d+1}$ with planar conformal boundary, whose metric in Kruskal coordinates is
\begin{align}\label{kruskalbh}
ds_{\rm BH}^2 = -A(UV) dU dV + B(UV) dx^2,
\end{align}
with $x \in \mathbb{R}^{d-1}$ the transverse direction, and we normalize $A(0) = B(0) = 1$ at the bifurcation surface. The future horizon for the right boundary is $U=0$, $V>0$.

\subsection{OTOC}\label{subsec:OTOCs}

The OTOC is a finite-temperature four-point function 
\begin{align}
C_\beta(t,x) &= \langle [V(0,x), W(t, 0)]^{\dagger} [V(0,x), W(t,0)] \rangle_{\beta},
\end{align}
where $V$, $W$ are local operators. In many chaotic systems, including holographic CFTs in the gravity limit, this correlator behaves at sufficiently large $t$ and $|x|$ as 
\begin{align}\label{otocformat}
C_\beta(t,x) \sim \exp\left[\lambda_L(t - t_* - |x| / \vbo ) \right].
\end{align}
When the OTOC takes this form, it defines the Lyapunov exponent $\lambda_L$, the scrambling time $t_*$, and the butterfly velocity $\vbo$. The interpretation is that if a perturbation is inserted at the origin, then the effects of chaos are large inside the butterfly cone $t > t_* + |x|/v_B$ \cite{Shenker:2013pqa,Shenker:2013yza,Roberts:2014isa,Gu:2016oyy}. As mentioned in the introduction, different probes of chaos can potentially see different butterfly cones, and $\vbo$ is defined as the butterfly velocity measured by the OTOC.

In the bulk, the insertion of the operator $W$ at large $t$ creates a shockwave on the horizon $U=0$, and the OTOC is calculated by treating $V$ as a probe that propagates through the shockwave. The shockwave metric takes the form
\begin{align} \label{eq:shock_metric}
ds^2_{\rm shock} &= ds^2_{\rm BH} + \delta(U) G(x) dU^2.
\end{align}
With this ansatz, the only nontrivial equation of motion in higher-derivative gravity is the component $E^{V}_U$. Away from sources, this component leads to the equation of motion \cite{Shenker:2013pqa,Roberts:2014isa,Shenker:2014cwa,Mezei:2016wfz,Dong:2022ucb}
\begin{align}\label{eq:shock_eq}
\left(-\p^2 + \frac{d(d-1)}{2} + \sum_{m=0}^{m_{\rm max}} a_m (\p^2)^m \right) G(x) = 0, 
\end{align}
where $\p^2=\delta^{ij} \p_i \p_j$ is the transverse Laplacian. This equation determines the transverse profile of a shockwave on the black hole horizon. The first two terms are the contributions from the Einstein action; the $a_m$'s are the contribution from higher-derivative corrections to the bulk theory, which in general are functions of the higher-derivative couplings and the background fields evaluated on the horizon. In $f(\mbox{Riemann})$ gravity, $m_\text{max} \leq 2$, while theories involving covariant derivatives acting on the Riemann tensors can have higher orders in transverse derivatives.

To discuss solutions of \eqref{eq:shock_eq} we must specify the sources and the boundary conditions at $|x| \to \infty$. A localized shockwave, which is the solution dual to a local operator insertion $W(t,0)$, has a delta-function source  at $x=0$ and satisfies the boundary condition $G(x) \to 0$ as $|x| \to \infty$. This leads to solutions which at large $|x|$ behave as
\begin{align}
G(x) \sim \frac{\mbox{const}}{|x|^a}e^{-\mu |x|}  \qquad \mbox{(local shock)},
\end{align}
for some $a$, and $\mu > 0$ is a root of the polynomial obtained by setting $\p^2 \to \mu^2$ in the differential operator in \eqref{eq:shock_eq}. If there is more than one positive root, we pick the one continuously connected to the Einstein gravity solution.
When this is translated into the boundary OTOC and compared to \eqref{otocformat}, it leads to the butterfly velocity \cite{Shenker:2013pqa,Roberts:2014isa,Shenker:2014cwa,Mezei:2016wfz,Dong:2022ucb}
\begin{align}
\vbo = \frac{\lambda_L}{\mu} = \frac{2\pi}{\beta \mu}.
\end{align}
We can consider other types of shockwaves by changing the sources. The shockwave directly relevant to the replica manifold discussed below is the solution
\begin{align}\label{planarshock}
G(x) = \mbox{const}\times  e^{\mu x^1} \qquad \mbox{(planar shock)},
\end{align}
where $x^1$ is one of the transverse directions. Since the shockwave equation \eqref{eq:shock_eq} has an even number of derivatives, it is clear that this solution has the same $\mu$ and therefore the same butterfly velocity as the local shock. It has a source at $x^1=+\infty$ that is extended in the other transverse directions $x^{2,3,\dots, d-1}$, and thus it appears in the calculation of the OTOC when $W$ is a nonlocal operator of codimension-two in the boundary spacetime.

\subsection{Pole skipping}\label{subsec:ps}
Pole skipping \cite{Grozdanov:2017ajz,Blake:2017ris,Blake:2018leo} occurs when the retarded thermal two-point function of the energy density, $\langle T_{00}T_{00}\rangle^{\rm ret}_\beta$, is ill-defined at a particular complex frequency $\omega_*$ and momentum $k_*$ (see also \cite{Grozdanov:2018kkt,Natsuume:2019sfp,Blake:2019otz,Grozdanov_2019,Natsuume_2020,Natsuume:2019vcv,Ahn:2019rnq,wu:2019esr,Choi:2020tdj,Ramirez:2020qer,Blake_2021,Grozdanov_2021,Blake:2021hjj,Yadav:2023hyg,Baishya:2023mgz,Wang:2022mcq,Ning:2023ggs,Loganayagam_2023,Grozdanov_2023,Grozdanov:2023e,Ahn_2024,Baishya:2024sym,Natsuume:2023nonbh,Natsuume:2023miss}). In the language of linear response, it arises when both the response and the source vanish at some energy and momentum, so that the correlator approaches $0/0$ as $(\omega, k_i) \to (\omega_*, k_*\hat{n})$. Here, $\hat{n}$ is a constant spacelike unit vector in the boundary that we choose to point in the $-x^1$ direction. The pole is `skipped' in the sense that a would-be pole of the retarded Green's function is canceled by a zero in the numerator.

This phenomenon is a harbinger of maximal chaos \cite{Blake:2017ris,Blake:2018leo}. The leading skipped pole is related to the butterfly effect parameters by 
\begin{align}\label{eq:ps_pt}
\omega_* = i \lambda_L = \frac{2\pi i}{\b}, \quad k_* =  \frac{i \lambda_L}{\vbp},
\end{align}
which defines the pole-skipping butterfly velocity, $\vbp$.

In the gravity limit of holographic theories, the retarded correlator is calculated by solving the linearized equations of motion in a black hole background, taking the form of linear wave equations with ingoing boundary conditions at the horizon. The advanced correlator is similar but with outgoing boundary conditions at the past horizon instead. The metric for a general stationary planar black hole can be written in ingoing Eddington-Finkelstein coordinates as
\be\label{eq:EFbh}
ds^2 =- f(r)dv^2+ 2dvdr  + h(r)dx^idx^i,
\ee
where $f(r_0)=0$ at the horizon $r = r_0$ and $x^i$, $i = 1, \dots, d-1$, are coordinates on $\mathbb{R}^{d-1}$. The inverse temperature is $\beta = 4\pi/f'(r_0)$. For simplicity, we assume that background matter fields are stationary, isotropic and homogeneous in $x^i$, in addition to being regular at both past and future horizons.

To study pole skipping, we consider linearized perturbations of dynamical fields around the stationary background, expanded in Fourier modes. Our focus is on the leading pole-skipping mode of metric perturbations,
\be\label{eq:deltagps}
\d g_{\m\n}(v,r,x) = \d g_{\m\n}(r) e^{-i\omega v+i k_1 x^1},
\ee
which is related to the chaos exponents, given by \eqref{eq:ps_pt}. The energy density two-point function is encoded in the component $\d g_{vv}$, which couples to the other components $\d g_{vr}$, $\d g_{vi}$, $\d g_{rr}$, \dots. In general, matter perturbations need to be turned on as well, but for our purposes they will not play an important role as long as we assume regularity of their background values at the horizon. Near the AdS boundary, the solution has a normalizable and non-normalizable branch. While the quasinormal modes are given by solutions with vanishing non-normalizable branch, the pole-skipping mode has a vanishing normalizable mode as well. In other words, the asymptotic boundary conditions are unperturbed.

Although the definition of pole-skipping involves the near-boundary behavior of the fields, in many theories including Einstein gravity \cite{Blake:2018leo} and its higher-derivative generalizations \cite{Grozdanov:2018kkt,Natsuume:2019sfp,Natsuume:2019vcv,wu:2019esr,Wang:2022mcq,Ning:2023ggs,Yadav:2023hyg,Baishya:2023mgz} it has been shown that the chaotic properties of the pole-skipping mode can be determined entirely from the leading near-horizon behavior. Furthermore, the pole-skipping mode has been identified as a certain horizon symmetry \cite{Knysh:2024asf} in maximally chaotic theories with an effective hydrodynamic description \cite{Blake:2017ris,Blake_2021}. We will therefore make the same assumption as in \cite{Wang:2022mcq,Ning:2023ggs}, namely that this general structure holds in the theory under consideration. In this case it is sufficient to expand the metric perturbations near the horizon in non-negative integer powers of $(r-r_0)$ at the pole-skipping point, $(\omega,k_1) = (\omega_*,-k_*)$:
\be\label{eq:deltagmn}
\delta g_{\m\n}(r) = \sum_{n=0}^{\infty} \delta g_{\m\n}^{(n)}(r-r_0)^n.
\ee
By choosing radial gauge, $\delta g_{r\m} = 0$, we can restrict to turning on only the modes $\delta g_{vv}$, $\delta g_{vi}$, and $\delta g_{ij}$ (and any matter fields). At the horizon, the leading non-trivial equation of motion is the component $E_{vv}$, and $\d g_{vv}^{(0)}$ decouples from the other components at the special point $\w = \w_*$, reducing to the homogeneous shockwave equation (cf. \eqref{eq:shock_eq}) \cite{Grozdanov:2017ajz,Blake:2018leo,Grozdanov:2018kkt,Ahn:2019rnq,Blake:2021hjj,Dong:2022ucb,Wang:2022mcq,Ning:2023ggs,Baishya:2024sym,wu:2019esr}
\be
\left(-\p^2 + \frac{d(d-1)}{2} + \sum_{m=0}^{m_{\rm max}} a_m (\p^2)^m \right) \d g_{vv}^{(0)}(x) = 0,
\ee
where $\d g_{\mu\nu}^{(0)}(x) \equiv \d g_{\mu\nu}^{(0)}e^{i k_1 x^1}$. The planar mode $\d g_{vv}^{(0)}(x)$ with $k_1 = -k_*$ solves this equation, from which one can extract the butterfly velocity
\be
\vbp = \frac{i\lambda_L}{k_*} = \frac{2\pi i}{\beta k_*}.
\ee
Higher order terms in the near-horizon expansion can in principle be solved order-by-order. In short, at the pole-skipping point, we have one less constraint on the metric perturbations from the equations of motion, providing an explanation of the universal ill-defined behavior of the retarded correlator.

As a concrete example, consider Einstein gravity. The $vv$-component of the Einstein equations at leading order is given by 
\be
\left(-\pa^2 - i \fr{d-1}{2}\omega h'(r_0) \right) \d g_{vv}^{(0)}(x) + \left(\omega - 2\pi i/\b\right)\delta^{ij}\left(\omega \delta g_{ij}^{(0)}(x) - 2  i \pa_i \d g_{vj}^{(0)}(x) \right)= 0.
\ee
If $\w$ is away from $\w_* = 2\pi i/\b$, this equation imposes a nontrivial relation between $\d g_{vv}^{(0)}(x)$, $\d g_{vi}^{(0)}(x)$, and $\d g_{ii}^{(0)}(x)$. However, at $\w = \w_*$, it reduces to
\be
\left(\pa^2 + i \fr{d-1}{2} \omega_* h'(r_0) \right) \d g_{vv}^{(0)}(x) = 0,
\ee
which is the homogeneous shockwave equation in Einstein gravity. Solving this equation for the planar mode indeed gives us the correct butterfly velocity.

\subsection{Entanglement wedge reconstruction}\label{subsec:ew}

In a holographic CFT, a boundary subregion $A$ is dual to the entanglement wedge (EW) of $A$, which is a region in the bulk defined as follows. Let $\gamma_A$ be the holographic entanglement surface, {\it i.e.}, the codimension-two surface in the bulk homologous to $A$ that extremizes the generalized area functional of higher-derivative gravity \cite{Ryu:2006bv,Ryu:2006ef, Hubeny:2007xt,Dong:2013qoa,Camps:2013zua,Dong:2019piw,Dong_2025}. The entanglement wedge of $A$ is the causal development of the bulk region enclosed by $\gamma_A \cup A$. (If there are multiple extrema then the entanglement wedge is defined by the one with minimal generalized area, but this will not be relevant here.)

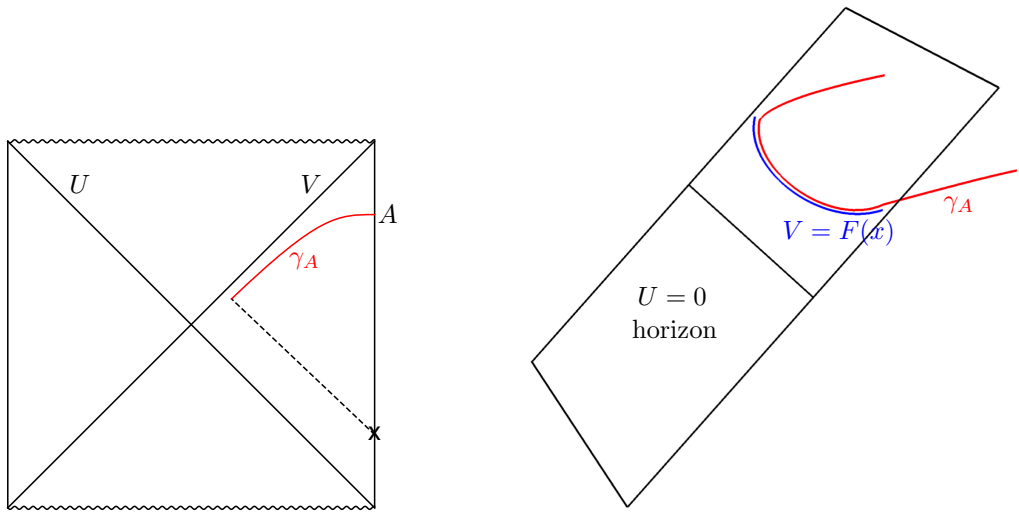
\begin{figure}[h]
\begin{center}
\begin{minipage}[b]{0.45\linewidth}
\flushleft
\begin{overpic}[scale=0.4]{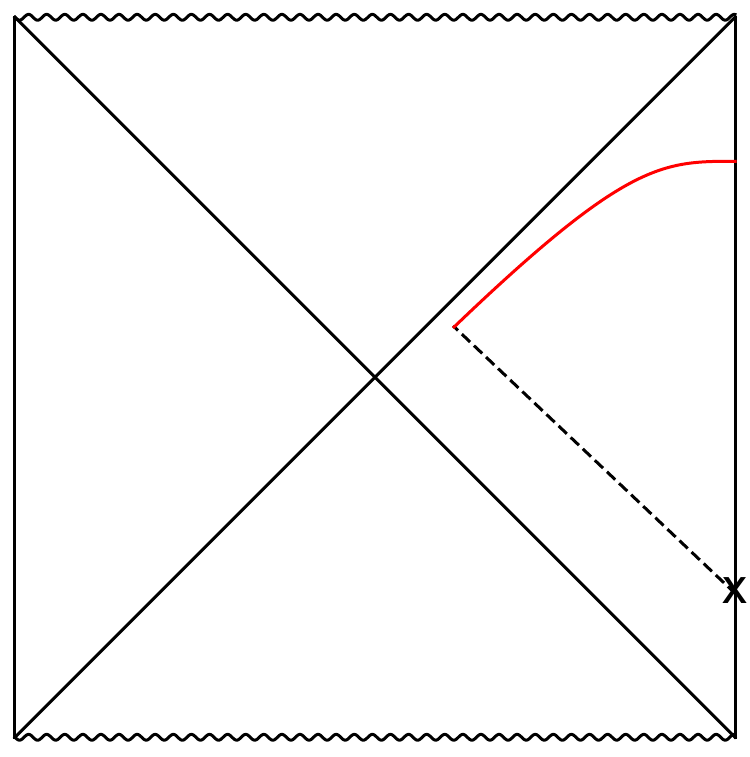}
\put(18,85){\parbox{0.2\linewidth}{$U $}}
\put(78,85){\parbox{0.2\linewidth}{$V $}}
\put(98,77){\parbox{0.2\linewidth}{$A $}}
\put(75,65){\parbox{0.2\linewidth}{\textcolor{red}{$\gamma_A $}}}
\end{overpic} 
\end{minipage}
\begin{minipage}[b]{0.45\linewidth}
\flushright
\begin{overpic}[scale=0.33]{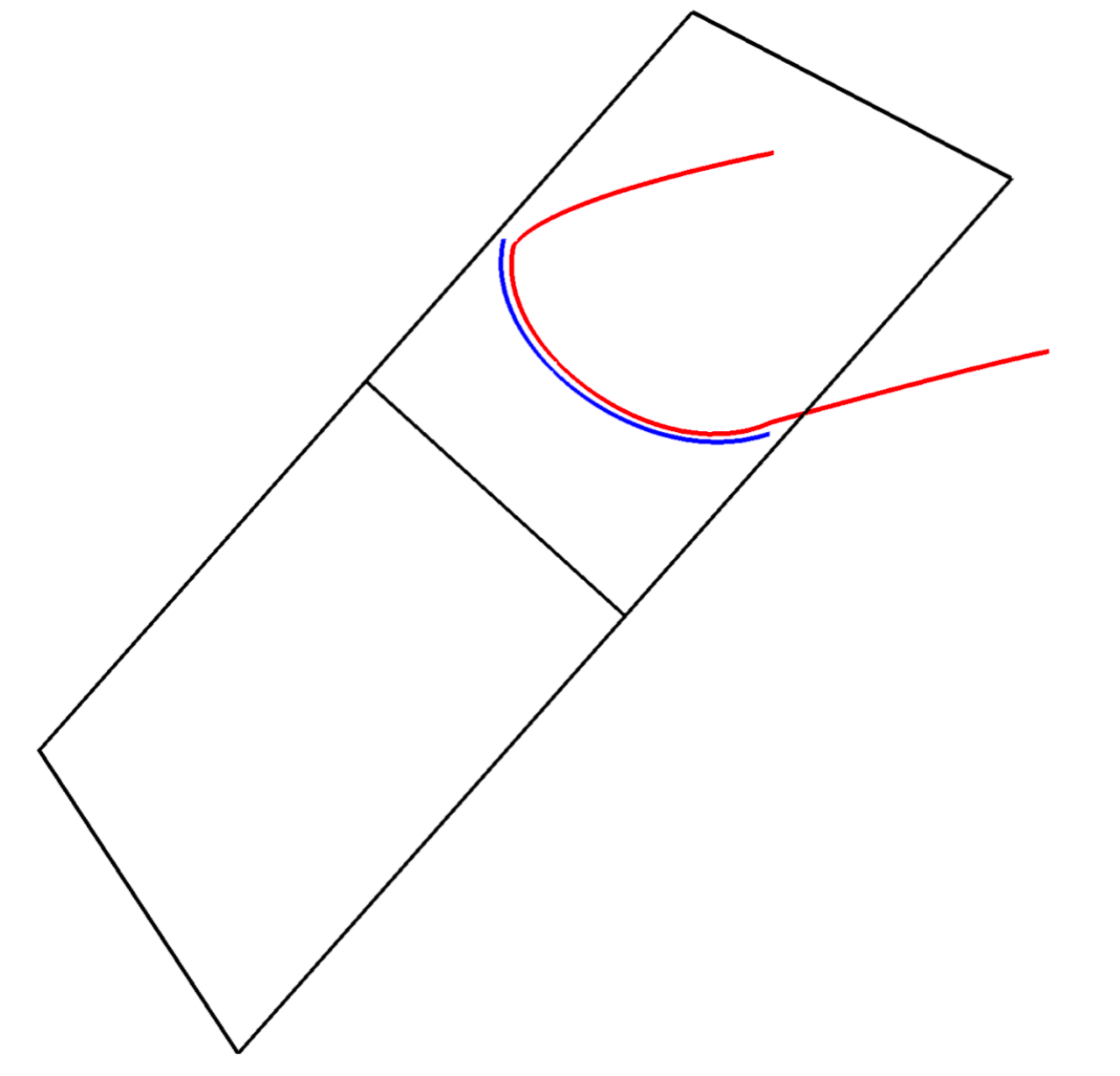}
\put(23,40){\parbox{0.2\linewidth}{$U = 0$}}
\put(22,34){\parbox{0.2\linewidth}{horizon}}
\put(50,52){\parbox{0.4\linewidth}{\textcolor{blue}{$V = F(x)$}}}
\put(80,57){\parbox{0.4\linewidth}{\textcolor{red}{$\gamma_A$}}}
\end{overpic}  
  \end{minipage}
\end{center}
    \caption{\small
    \textit{Left:} To define the butterfly velocity from entanglement wedge reconstruction, a particle is dropped into the black hole and falls on the dashed line. The particle can be reconstructed in region $A$ if the entanglement wedge bounded by $\gamma_A \cup A$ contains the particle.
    \textit{Right:} At late times, taking the region size $R$ and time $T$ large with $\vbe$ held fixed, the entanglement surface (red) approaches a special extremal surface (blue) exactly on the future horizon.\label{fig:EW_spherical}
    }
\end{figure}

The third notion of the butterfly velocity $\vbe$ is the speed at which a local operator ${\cal O}$ grows in size, as measured by the entanglement wedge needed to reconstruct the operator at time $T$, ${\cal O}(T) = e^{iHT} {\cal O}(0)e^{-iHT}$ \cite{Mezei:2016wfz}. To calculate this velocity in a thermal state at inverse temperature $\beta$, we suppose that the operator ${\cal O}(0)$ creates a highly energetic probe particle that falls into a black hole on a (nearly-)null geodesic. The minimal region $A(T)$ needed to reconstruct the operator ${\cal O}(T)$ is a ball with some radius $R$ whose entanglement surface $\gamma_{A(T)}$ touches the infalling particle at its tip. This is illustrated in figure \ref{fig:EW_spherical}. At late times, the infalling particle is near the horizon, the operator size grows linearly by this measure, and this defines the third butterfly velocity $\vbe = R/T$. In $d=2$ boundary dimensions, one finds $\vbe=1$, while for $d>2$ the entanglement surface extends into the bulk further than the causal wedge, which results in $\vbe < 1$.

To calculate $\vbe$ it is convenient to translate the whole experiment back in time so that the particle is dropped at time $t=-T$, where $t$ is the Schwarzschild time coordinate, and consider the entanglement wedge at time zero. We will boost back to the original experiment at the end. For a region $A$ at $t=0$ on the boundary, the entanglement surface is specified in Kruskal coordinates \eqref{kruskalbh} by
\begin{align}
U = -F(x) , \quad V = F(x). 
\end{align}
For a ball-shaped region centered at the origin, $F$ is a function only of $|x|$. The infalling particle at time zero is at 
\begin{align}
V_0 = - U_0 \sim e^{-\lambda_L T},
\end{align}
up to an ${\cal O}(1)$ constant.
Therefore we set $F(0) = V_0$, so that the tip of the entanglement surface meets the particle, then solve the (higher-derivative) extremal surface equation for $F(x)$. One finds that for large enough $T$, the extremal surface is in the near-horizon region $|U|, |V| \ll 1$, and it has an exponential profile in the transverse directions when $|x| \gg \beta$: 
\begin{align}\label{sphericalF}
F(x) = \frac{\text{const}}{|x|^a}e^{-\lambda_L T + \nu |x|}.
\end{align}
At a radius of order $|x| \sim \lambda_L T / \nu$, the extremal surface exits the near-horizon region and heads toward the boundary. Since we only care about the exponential dependence on $|x|$, we can ignore that part of the profile for the purposes of determining where it lands at the boundary, and approximate the landing radius as $\lambda_L T/\nu$. The conclusion is that the region whose entanglement wedge marginally contains the infalling particle has radius 
\begin{align}\label{vbfromnu}
R(T) = \vbe T , \qquad \vbe = \frac{\lambda_L}{\nu}. 
\end{align}
This is the holographic result for the third butterfly velocity.

This calculation was carried out in a general four-derivative theory of gravity in \cite{Mezei:2016wfz}, and it was observed that the extremal surface equation for $F(x)$ is identical to the homogeneous shockwave equation \eqref{eq:shock_eq}. It follows that $\vbe = \vbo$ in these theories. The calculation was extended to arbitrary $f(\mbox{Riemann})$ theories in \cite{Dong:2022ucb}, and once again, the extremal surface and homogeneous shockwave equations turn out to be identical. One example involving $\del_\mu R$ was also verified in \cite{Dong:2022ucb}, suggesting that the equality is a general feature of holographic theories in the gravity limit --- or perhaps all maximally chaotic quantum theories, with some appropriate generalization of entanglement wedge reconstruction. As emphasized in \cite{Dong:2022ucb} and reviewed in the introduction, the agreement between these two butterfly velocities is not obvious. The two calculations leading to the equations for $G(x)$ (the shockwave) and $F(x)$ (the extremal surface) seem very different, and it is only by analyzing each one explicitly that they have been shown to agree.

\subsubsection{Horizon limit}

Let us now return to the original frame, where the particle is dropped at time $t=0$ and the extremal surface is in the slice $t=T$. This is related to the previous frame by the time translation $t \to t + T$, which in the bulk translates to a boost $U \to e^{-\lambda_L T} U$, $V \to e^{\lambda_L T} V$. The extremal surface now sits at
\begin{equation}
    U = -e^{-\lambda_L T} F(x), \qu V = e^{\lambda_L T} F(x).
\end{equation}
Consider the entanglement wedge whose tip follows the infalling particle all the way to the horizon, $U \to 0$. On the boundary, this corresponds to taking the limit $R$, $T \to \infty$, holding $R/T = \vbe$ fixed. For large $T$ the near-horizon portion of the surface is very close to the horizon, and in the limit $T \to \infty$ it approaches a special extremal surface that is exactly at $U=0$. See figure \ref{fig:EW_spherical}. The special limiting surface is the curve
\begin{align}\label{eq:UVf}
U = 0, \qquad V = F(x),
\end{align}
with $F(x) \sim e^{\nu |x|}$ at large $|x|$.\footnote{Note that we have rescaled $F$ by the boost factor so that it remains finite as $T\to\infty$, which is the bulk version of holding $R/T$ fixed.} This limiting surface is similar to the one discussed in \cite{Hartman:2013qma} in the context of extremal surfaces that pass through the black hole interior (commonly known as the Hartman-Maldacena surface), and to the holographic entanglement membrane developed in \cite{Mezei:2016zxg,Jonay:2018yei,Mezei:2018jco,Mezei_2020}. It is not the holographic entanglement surface for any boundary region because it remains on the horizon (except in three bulk dimensions, where it reaches the boundary), but it controls the universal dynamics of the entanglement entropy and entanglement wedge for boundary regions in the late-time limit. The butterfly velocity $\vbe$ can be extracted from this surface using \eqref{vbfromnu}.

\subsubsection{Planar version}
The entanglement-wedge butterfly velocity can also be extracted from a calculation with planar symmetry \cite{Mezei:2016zxg,Mezei:2018jco}.  This will turn out to be the entanglement calculation that is most directly connected to pole skipping. Consider a region $A$ which is a half-space in the CFT on the right side of the eternal black hole,
\begin{align}\label{eq:half_space}
    A =  \{ t = T, \ x^1 < v T \mbox{\ in the right CFT} \}.
\end{align}
The bulk entanglement surface $\gamma_A$ approaches very close to the horizon as $x^1 \to -\infty$, behaving as 
\begin{align}
    U \sim  -e^{-\lambda_L T +\nu(x^1 - vT)} , \qquad V \sim e^{\lambda_L T + \nu(x^1-vT)},
\end{align}
with the same exponent $\nu$ as in \eqref{sphericalF}. If we set $v = \vbe$ and take $T \to \infty$, then $\gamma_A$ approaches the limiting extremal surface
\begin{align}\label{eq:planar_UVf}
    U = 0 , \qquad V = F(x) ,\qquad \mbox{\ with \ } F(x) = \mbox{const}\times e^{\nu x^1}
\end{align}
on the future horizon.  See figure \ref{fig:btz} below for an illustration of this extremal surface in three-dimensional gravity.

This result is similar to \eqref{eq:UVf} except that now $F(x)$ has planar symmetry.  In both cases, the differential equation satisfied by $F(x)$ is the same --- only the boundary conditions differ --- so the exponent $\nu$ and the butterfly velocity $\vbe$ inferred from it are identical.

\section{Pole skipping  = shockwave}\label{sec:psshock}

In this section we will review the result of \cite{Wang:2022mcq,Ning:2023ggs} showing the equality $\vbp = \vbo$ in higher-derivative gravity, and in addition, discuss the pole-skipping mode away from the near-horizon region. We will assume that the pole-skipping mode for metric perturbations exists and that it can be expanded in non-negative integer powers of $(r-r_0)$ away from the horizon; this is true in many examples \cite{Grozdanov:2017ajz,Blake:2018leo,Blake:2019otz,Grozdanov:2018kkt,Grozdanov_2019,Natsuume:2019sfp,Natsuume_2020,Natsuume:2019vcv,wu:2019esr,Ahn:2019rnq,Blake:2021hjj,Yadav:2023hyg,Baishya:2023mgz,Baishya:2024sym,Loganayagam_2023,Grozdanov_2023,Ahn_2024,Natsuume:2023nonbh,Natsuume:2023miss} but it has not been established whether this is universal or applies only to certain bulk gravity theories. Furthermore, we will assume that this mode is the leading one in the near-horizon expansion; as argued in \cite{Wang:2022mcq,Ning:2023ggs}, this is true as long as the gravity theory only has minimally-coupled matter fields of spin less than two. We also give a slight variant of the argument in \cite{Wang:2022mcq,Ning:2023ggs} based on identifying the shockwave as the imaginary part of the pole-skipping mode.

Let us write the metric perturbations as 
\be
\delta g_{\m\n}(v,r,x) = \delta g_{\m\n}(r) G(x) e^{-i\omega v}, \qu ~~G(x) = e^{i k_1 x^1},
\ee
where $\delta g_{\m\n}(r)$ is assumed to have a Taylor expansion at $r = r_0$ and we again impose radial gauge $\d g_{\m r}=0$. We can write an ansatz for the full pole-skipping solution
\be
ds^2_{\rm PS} = ds^2_{\rm BH} + G(x) e^{-i\w v} \Big(\d g_{vv}(r) dv^2 + \d g_{vi}(r)dvdx^i + \d g_{ij}(r)dx^idx^j\Big),
\ee
which solves the linearized equations of motion everywhere with vanishing normalizable and non-normalizable perturbations at the boundary. As reviewed in section \ref{subsec:ps}, pole skipping occurs when the equation of motion for $\delta g_{vv}$ decouples at the horizon, and this determines the pole-skipping point $(\omega, k_1) = (\omega_*, -k_*)$.

It turns out for our purposes that it is more convenient to write the ansatz in Kruskal coordinates. This will in turn make its relation to the shockwave \eqref{eq:shock_metric}, as well as regularity at the horizon, manifest. The coordinate change covering the right exterior is
\begin{align}\label{eq:kruskal2ef}
U = -e^{-\fr{f'(r_0)}{2}(v-2r_*)}, \qu V = e^{\fr{f'(r_0)}{2}v},
\end{align}
where $dr_*/dr = 1/f(r)$ defines the tortoise coordinate $r_*$. The stationary black hole in these coordinates is given by \eqref{kruskalbh} where
\be\label{eq:Kruskal2sch}
A(UV) =-\fr{4f(r)}{f'(r_0)^2 UV}, \qu B(UV) = h(r).
\ee
The future and past horizons sit at $U = 0$, $V > 0$ and $V = 0$, $U < 0$, respectively; the bifurcation surface is at $U=V = 0$ and the AdS boundary is at $UV = -1$. In these coordinates, the ingoing mode becomes $e^{-i\omega v} = V^{-2i\omega/f’(r_0)}$ and therefore the metric perturbation is given by
\begin{equation}\label{eq:ps_mode0}
\delta ds^2 = G(x) V^{-2i\omega/f’(r_0)} \bigg(H_{VV}(UV) \frac{dV^2}{V^2} + H_{Vi}(UV) \frac{dV}{V} dx^i + H_{ij}(UV)dx^i dx^j\bigg),
\end{equation}
where we have defined the functions $H_{VV}(UV) \propto \delta g_{vv}(r)$, $H_{Vi}(UV) \propto \delta g_{vi}(r)$ and $H_{ij}(UV) = \delta g_{ij}(r)$, with $r$ given implicitly by \eqref{eq:Kruskal2sch}. Since the functions $\delta g_{\m\n}(r)$ have Taylor expansions at $r = r_0$ and $UV \sim (r-r_0) + \mo((r-r_0)^2)$ near the horizon, the functions $H_{\m\n}$ also have Taylor expansions at $V = 0$:
\begin{equation}
H_{\m\n}(UV) = \sum_{n=0}^{\infty} H_{\m\n}^{(n)} (U V)^n, \qu \m,\n = \{V,i\}.
\end{equation}
It is easy to see that for generic values of $\w$, the metric \eqref{eq:ps_mode0} is not regular at the horizon when $H_{VV}^{(0)}$ is nonzero. However, at $\w = \w_*$, we have
\begin{equation}\label{eq:ps_mode}
\delta ds^2_\text{PS} = G(x) \bigg(H_{VV}(UV) \frac{dV^2}{V} + H_{Vi}(UV) dVdx^i + V H_{ij}(UV)dx^i dx^j\bigg).
\end{equation}
Now that the expansion has integer powers there is no longer a branch cut at $V=0$, and in fact the perturbation is regular on both the past and future horizons \cite{Natsuume:2019sfp}.

So far, our discussion has focused on the ingoing mode, which gives the retarded correlator on the boundary. However, the outgoing mode, which gives the advanced correlator, and in general is regular (singular) at the past (future) horizon, is more convenient for the application below where we study entanglement wedges at late times as opposed to early times. This amounts to exchanging $V$ and $U$ in the pole-skipping metric. Therefore, adding the advanced pole-skipping mode to the black hole background we have the following spacetime, recorded for later comparison to the replica manifold:
\begin{equation}\label{eq:ps_ansatz}
    ds^2_\text{PS} = ds^2_{\text{BH}} + G(x) \bigg(H_{UU}(UV) \frac{dU^2}{U} +  H_{Ui}(U V) dUdx^i+ U H_{ij}(UV) dx^idx^j  \bigg).
\end{equation}
where, as above, the functions $H_{\m\n}$ have expansions in non-integer powers in $UV$ with $H_{UU}^{(0)}$ nonzero. From now on, we will set $H_{UU}^{(0)} = 1$ for simplicity; it can be restored by an appropriate rescaling of $G(x)$.

To understand why the equation of motion for the pole-skipping mode reduces to the homogeneous shockwave equation, let us perform the infinitesimal shift $U \to U - i\varepsilon$. This regulates the $UU$-component of the metric at the horizon $U = 0$,
which at leading-order in the near-horizon expansion goes as
\be
\d g_{UU} = G(x) \fr{1}{U} + \mathcal{O}(U^0),
\ee
while \eqref{eq:ps_ansatz} remains a solution to the linearized equations of motion. In the regular terms, the shift has no effect, so in those terms we can set $\varepsilon = 0$. Hence the pole-skipping metric takes the form
\be
ds^2_{\rm PS} = -A(UV)dUdV+B(UV)dx^2 + G(x) \, \frac{dU^2}{U-i\varepsilon} + \cdots.
\ee
where the dots denote terms regular at $U=0$. Using ${\rm Im} \, \frac{1}{U-i\varepsilon} = \pi \delta(U)$, we see that the imaginary part of the perturbation is a shockwave localized at the horizon,
\be
{\rm Im} \, \d ds^2_{\rm PS} = \pi \d ds^2_{\rm shock}.
\ee
The subleading terms away from the horizon are purely real. Since the real and imaginary parts of a linearized perturbation must each satisfy the equations of motion individually, it follows that $G(x)$ is the shockwave profile \eqref{planarshock}. This establishes the equality of the two corresponding butterfly velocities \cite{Grozdanov:2017ajz,Blake:2018leo,Grozdanov:2018kkt,Dong:2022ucb,Wang:2022mcq,Ning:2023ggs,Baishya:2024sym,Ahn:2019rnq,Grozdanov_2023,Grozdanov_2021},
\begin{align}
\vbp = \vbo,
\end{align}
in general theories of gravity for which the pole-skipping solution \eqref{eq:ps_ansatz} exists.

\section{Pole skipping = replica manifold}\label{sec:ps_equals_displacement}
In this section, we construct gravitational replica manifolds \cite{Lewkowycz:2013nqa,Dong:2013qoa,Camps:2013zua,Dong:2016fnf,Dong:2017xht,Dong:2019piw,Colin-Ellerin:2020mva,Colin-Ellerin:2021jev,Dong_2025} branched along a cross section of a black hole horizon, at leading order in the replica limit $n \to 1$. As reviewed in section \ref{subsec:ew}, the late-time entanglement wedge of a boundary region $A$ at finite temperature is controlled by an extremal surface that is exactly on the horizon. The late-time limit is defined by scaling the time $T$ and the region size $R$ to infinity along the butterfly cone, with the ratio $R/T = \vbe$ held fixed. Therefore, the replica manifold considered here is what determines the late-time entanglement entropy, entanglement wedge, and butterfly velocity $\vbe$.

We will show that when the extremal surface is near the bifurcation surface of the black hole, the replica manifold to leading order in $(n-1)$ is given by the black hole plus the pole-skipping mode. To be more precise, the replica manifold, upon analytically continuing to Lorentzian signature, is the metric \eqref{eq:ps_ansatz} upon identifying the extremal surface profile $F$ with a particular function $G$ satisfying the homogeneous shockwave equation. The standard pole-skipping metric, with linear momentum in the $x^1$-direction, corresponds to the solution $G(x)=(n-1)F(x)\sim e^{\nu x^1}$,  and thus to the entanglement wedge with planar symmetry discussed around \eqref{eq:planar_UVf}. This is the main result of the paper.

\subsection{Formulating the problem}
In the replica method, the entanglement entropy of the state $\rho$ is obtained from the replica partition function $Z_n = \mbox{tr}\,\rho^n$ as
\begin{align}\label{eq:ee_replica}
S = -\partial_n Z_n|_{n=1} = \lim_{n \to 1} \frac{1}{1-n} \log Z_n.
\end{align}
In holographic theories, the replica partition function is calculated using the gravitational path integral. The boundary conditions are set by the replicated system $\rho^{\otimes n}$, which has a $\mathbb{Z}_n$ symmetry permuting the replicas. The replica partition function is evaluated on the dominant saddle point which we call the replica manifold ${\cal M}_n$. Following \cite{Lewkowycz:2013nqa,Dong:2013qoa,Camps:2013zua,Dong:2016fnf,Dong:2017xht,Colin-Ellerin:2020mva,Colin-Ellerin:2021jev,Dong:2019piw,Dong_2025}, ${\cal M}_n$ is assumed to be replica symmetric, so that the quotient
\begin{align}
\widetilde{\cal M}_n  = {\cal M}_n / \mathbb{Z}_n
\end{align}
is well-defined. The full replica geometry ${\cal M}_n$ must satisfy the (higher-derivative) equations of motion, while the quotient $\widetilde{\cal M}_n$ has a $2\pi/n$ conical defect at the branching surface. In general, the replica manifold is hard to solve for explicitly except in some simple cases, but in the replica limit $n \to 1$ the problem simplifies and admits a general solution.

Let us now specialize to the replica manifold for the late-time entanglement wedge of the half-space region \eqref{eq:half_space}. Our goal is to explicitly construct the geometry to leading order in $(n-1)$, from which we extract the dynamics of the entanglement entropy and entanglement wedge (cf. \eqref{eq:ee_replica}). Consider a codimension-two extremal surface on the horizon of the Lorentzian black hole \eqref{kruskalbh}, specified by
\begin{align}
U = 0 \ , \quad V = F(x).
\end{align}
This surface, which can be understood as a null displacement from the bifurcation surface, is the limiting value of the late-time entanglement wedge which encodes the butterfly velocity $\vbe$.
The corresponding replica manifold ${\cal M}_n$ is a complex solution to the equations of motion that is branched around this surface. To describe it, we first shift the branch cut to the origin by performing the coordinate change $V = V' + F$. The Lorentzian black hole \eqref{kruskalbh} in these coordinates is
\begin{align}
ds^2 = -A( U(V'+F) )dU (dV' + \p_i F dx^i) + B(U(V'+F))dx^2, 
\end{align}
with extremal surface sitting at $U = V' = 0$. We now continue to Euclidean signature by introducing a complex coordinate $z \in \mathbb{C}$, and taking $U \to -z$, $V' \to \bz$ to obtain
\begin{align}\label{m1}
ds^2 = A(-z(\bz+F))dz (d\bz + \p_i F dx^i)  + B(-z(\bz+F)) dx^2. 
\end{align}
This is the background geometry ${\cal M}_1$. The metric is complex because we have chosen to branch around a nonstationary point.

The replica manifold ${\cal M}_n$ we wish to construct is replica symmetric and branched around the origin. In terms of complex coordinates, with $(z,\bz) = (0,0)$ the origin, this requires the metric to be invariant under
\begin{align}\label{replicashift}
(z, \bz) \to (z e^{2\pi i/n}, \bz e^{-2\pi i/n}). 
\end{align}
This implies the metric is single-valued on the quotient ${\cal \widetilde{M}}_n$. To take the limit $n\to 1$, we assume that the quotient metric remains single-valued as we continue to non-integer $n$. However, it turns out that the geometry relevant to us is the full replica manifold ${\cal M}_n$, taken near $n = 1$, instead. We therefore impose that the metric of ${\cal M}_n$ is invariant under \eqref{replicashift} also at non-integer $n$. 

In general, the geometry of the replica manifold ${\cal M}_n$ for noninteger $n$ is not smooth, and is therefore not a solution. As we will see, however, at least to leading order in the replica limit $n \to 1$, as well as the displacement $F$ away from the bifurcation surface, one can find a smooth solution to the linearized equations of motion. To summarize, our goal is to find a replica manifold ${\cal M}_n$ that is non-singular everywhere including the origin, is invariant under \eqref{replicashift}, and reduces to the background ${\cal M}_1$ at $n = 1$.

\subsection{Replica manifold from pole skipping}

The solution near the origin is provided by the advanced pole-skipping metric, as we will now demonstrate by an explicit change of coordinates. The advanced pole-skipping metric from \eqref{eq:ps_ansatz} is, at leading order in the near-horizon expansion,
\begin{align}\label{rps}
ds^2_{\rm PS} &= ds^2_{\rm BH} + \delta ds^2_{\rm PS} = -A(UV) dU dV  + B(UV) dx^2  + G(x) \bigg(\frac{dU^2}{U} + \cdots\bigg)
\end{align}
where, for now, $G$ is any solution to the homogeneous shockwave equation \eqref{eq:shock_eq}. The dots indicate corrections away from the horizon at $U=0$. As discussed in section \ref{sec:psshock}, these corrections can be expanded in non-negative integer powers of $U$, $V$. 

Let us now perform the coordinate change
\begin{align}
U = -z , \qquad V = \bz + \frac{z^{n-1}}{n-1} G,
\end{align}
which should be understood as the leading terms in an expansion around $z = 0$. We will work in the limit $n \to 1$, and keep up to $\mo(n-1)$ terms. To get a well-defined limit, we choose $G(x) = (n-1)F(x)$ with $F$ any $\mo((n-1)^0)$ solution to \eqref{eq:shock_eq}.
The pole-skipping metric \eqref{rps} becomes 
\begin{align}\label{rmf}
\begin{split}
 ds^2_{\rm PS} &= 
A(-z(\bz + z^{n-1}F)) dz (d\bz + z^{n-1} \p_i F dx^i)+ B(-z(\bz + z^{n-1}F)) dx^2 \\
&\qquad  + (n-1) F \times \mbox{(corrections away from $z=0$)} + {\cal O}((n-1)^2).
\end{split}
\end{align}
We claim this is the replica manifold ${\cal M}_n$ to leading order in $(n-1)$ and the displacement $F$, expanded around $z = 0$. Since we are working to $\mo(n-1)$, one can replace $z^{n-1} \to 1 + (n-1) \log z$ in this expression, but we have retained the power to make the replica transformation \eqref{replicashift} manifest.

The metric \eqref{rmf} satisfies all of the conditions laid out in the previous section: It is non-singular and solves the equations of motion at leading order near $z = 0$, reduces to the background ${\cal M}_1$ at $n=1$, and is invariant under \eqref{replicashift}. The latter property is easily checked for the first line in \eqref{rmf}, and it holds for the corrections on the second line because \eqref{replicashift} acts trivially at $\mo((n-1)^0)$, {\it i.e.}, $(z,\bz) \to (z,\bz) + \mo(n-1)$. Therefore this is the leading-order replica manifold near $z = 0$. To establish that this is the full replica manifold, we need to show that \eqref{rmf}, or equivalently \eqref{rps} for the Lorentzian manifold, can be extended away from $z = 0$ while retaining all of the above conditions. As we now show, the extension coincides with the full pole-skipping solution.

\subsection{Replica manifold away from the horizon}

We have shown that at $\mo(n-1)$ and at leading order in $U \to 0$ and the displacement $F$, the Lorentzian replica manifold takes the form of the pole-skipping mode at the future horizon,
\be \label{eq:replica_leading}
\delta ds^2_n \simeq (n-1)F(x) \bigg(\frac{dU^2}{U} + \dots\bigg),
\ee
where $F$ is any $\mo((n-1)^0)$ solution of the homogeneous shockwave equation. To complete the argument showing that these modes are in fact identical, we must show that the two agree even away from the horizon.

The background metric is invariant under boosts
\be
U \to \a U, \qu V \to \a^{-1} V,
\ee
which on the boundary correspond to time translations $t \to t + T$ for $T = \lambda_L^{-1} \log \a$. Under the condition that the late-time entanglement wedge follows the butterfly cone, {\it i.e.}, invariance under the shifts $t \to t + T$, $x^1 \to x^1 + \vbe T$ as $T \to \infty$ (an assumption that will be justified \textit{a posteriori} by finding such a solution), this implies that the replica manifold up to $\mo(n-1)$ is invariant under the following combination of boost plus rescaling of the extremal surface:
\be
U \to \a U, \qu V \to \a^{-1} V, \qu F = \a^{-1} F,
\ee
where $F(x) \propto e^{\nu x^1}$ with $\nu = \lambda_L/\vbe$. Note that this statement holds beyond linear order in $F$. As such, we can write an ansatz for the full nonlinear replica manifold at $\mo(n-1)$ as, in radial gauge, 
\be\label{eq:full_replica}
\delta ds_n^2 = (n-1) F(x) \bigg(\widetilde{H}_{UU}\big(UV, UF\big) \frac{dU^2}{U} + \widetilde{H}_{Ui}\big(UV, UF\big) dUdx^i + U \widetilde{H}_{ij}\big(UV, UF\big) dx^idx^j\bigg),
\ee
where the functions $\widetilde{H}_{\m\n}$ have Taylor expansions at $U = 0$ to ensure regularity at the horizon, and $\widetilde{H}_{UU} \simeq 1 + \mo(U)$ such that the leading behavior is given by \eqref{eq:replica_leading} to match the replica boundary conditions after continuing to Euclidean signature (cf. \eqref{rmf}). At the asymptotic boundary, the boundary conditions are unperturbed at this order in $(n-1)$. The reason is that the extremal surface, and therefore the branching surface, lies entirely on the horizon and never reaches the boundary (except in three dimensions, where it intersects the boundary at future null infinity).

To see the relation to the full pole-skipping mode, we expand the replica metric to linear order in $F$, which we can write as
\be\label{eq:rm_lin}
\delta ds_n^2 = (n-1) F(x) \bigg(\widetilde{H}_{UU,0}\big(UV\big) \frac{dU^2}{U} + \widetilde{H}_{Ui,0}\big(UV\big) dUdx^i + U \widetilde{H}_{ij,0}\big(UV\big) dx^idx^j\bigg),
\ee
where $\widetilde{H}_{\mu\nu,0}\big(UV\big) \equiv \widetilde{H}_{\mu\nu}\big(UV,0\big)$. This, by definition, is the displacement mode for the replica manifold branched around the bifurcate horizon, with null displacement along the horizon by $V \to V + F$. Eq. \eqref{eq:rm_lin} takes precisely the form of the pole-skipping metric we wrote down in \eqref{eq:ps_ansatz}, provided that we can identify
\begin{equation}
\widetilde{H}_{\mu\nu,0}(U V) = H_{\mu\nu}(U V), \qu \mu,\nu = \{U,i\}.
\end{equation}
Since we have assumed the same boundary conditions at both the horizon and asymptotic boundary, uniqueness of the bulk solution implies that the two solutions must be identical. In other words,
\be
\delta ds_n^2 = (n-1) \delta ds^2_{\rm PS}.
\ee
This establishes the equality of the two butterfly velocities,
\begin{align}\label{vb1}
\vbe = \vbp.
\end{align}
It is already known that the pole-skipping metric is related to the shockwave and $\vbo = \vbp$, as reviewed in section \ref{sec:psshock}, so we can chain these results together to conclude that
\be
{\rm Im} \, \delta ds_n^2 = \pi (n-1) \delta ds^2_{\rm shock}
\ee
and $\vbe = \vbo$.

\section{Example: Planar BTZ in Einstein gravity}\label{sec:BTZ}
To illustrate the general relationship between the replica manifold, displacement mode, and pole-skipping mode we will now work out the details for the planar BTZ black hole. We start by reviewing the construction of the replica manifold for null-deformed half space in the vacuum state. In other words, we describe the bulk dual of the Cardy-Calabrese calculation \cite{Calabrese:2009qy} of R\'{e}nyi entropy in two-dimensional CFT. This is related to the black hole by a diffeomorphism. This will allow us to find the linearized displacement mode exactly ({\it i.e.}, to linear order in $(n-1)$ but not limited to the near-horizon region). This result is a special case of the holographic dual of the R\'{e}nyi twist displacement operator found in \cite{Balakrishnan:2016ttg}, derived by a different method. We will also find the explicit coordinate change to the pole-skipping mode in radial gauge.

\subsection{Replica manifold in the vacuum state}

\subsubsection{Conformal transformations of AdS$_3$}
Consider a two-dimensional CFT in the vacuum state. The Euclidean theory lives on the complex plane, $z \in \mathbb{C}$, and the bulk geometry is Poincar\'e AdS$_3$, 
\begin{align}\label{poinz}
ds^2 = \frac{dz d \bz + d\xi^2}{\xi^2}.
\end{align}
The bulk dual of a conformal transformation $z = f(w)$, $\bz = f(\bw)$ in the CFT is a diffeomorphism that takes this form at the boundary while preserving  the AdS$_3$ boundary conditions.  In Fefferman-Graham gauge this requirement fixes the full coordinate change to  \cite{Roberts:2012aq}
\begin{align}\label{eq:roberts}
\begin{split}
z &= f(w) - \frac{2\zeta^2 (f')^2 \barf''}{4 f' \barf' + \zeta^2 f'' \barf''} \\
\bz &= \barf(\bw)  - \frac{2\zeta^2 f'' (\barf')^2}{4 f' \barf' + \zeta^2 f'' \barf''} \\
\xi &= \zeta \frac{4 (f' \barf')^{3/2}}{4 f' \barf' + \zeta^2 f'' \barf''}
\end{split}
\end{align}
The resulting metric is
\begin{align}\label{schmet}
ds^2 &= \frac{dw d\bw + d\zeta^2}{\zeta^2} - 
\frac{1}{2}\{f,w\}dw^2 - \frac{1}{2}\{\barf,\bw\} d\bw^2
 + \frac{\zeta^2}{4} \{f,w\} \{\barf,\bw\} dw d\bw
\end{align}
where $\{ f,w \} = -\frac{3}{2} \frac{(f'')^2}{(f')^2} + \frac{f'''}{f'}$ is the Schwarzian derivative. In this gauge, the Brown-York tensor is related to the $\mo(\zeta^0)$ components of the metric, denoted $g_{\mu\nu}^{(0)}$, by  $T_{ww} =  \frac{1}{8 \pi G} g_{ww}^{(0)}$, which agrees with the usual CFT expression $T_{ww} \!=\! -\frac{c}{24\pi} \{f,w\}$ with the Brown-Henneaux central charge, $c = \frac{3}{2G}$. 

\subsubsection{Gravity dual of Cardy-Calabrese}

Let region $A(b,\bb)$ be the interval with endpoints at $(0,0)$ and $(b,\bb)$. The replica partition function $\mbox{Tr}\, \rho_A^n$ is computed in the CFT by a path integral on an $n$-sheeted cover of the complex plane, branched along $A$ \cite{Calabrese:2009qy, Hung:2011nu}. Equivalently, it is the correlation function of order-$n$ twist operators inserted at the endpoints. We will view $w \in \mathbb{C}$ as the coordinate where the CFT is originally defined, and $z \in \mathbb{C}$ as the global coordinate on the cover.  They are related by the uniformizing map
\begin{align}
z = f(w) = \left( \frac{w}{b-w} \right)^{1/n}, \quad
\bz = \barf(\bw) = \left( \frac{\bw}{\bb - \bw} \right)^{1/n} . 
\end{align}
The bulk replica manifold is therefore \eqref{schmet}, with this particular choice of $f$ and $\barf$, which have
\begin{align}
\{f,w\} = \frac{(n^2-1)b^2}{2n^2w^2(b-w)^2},  \qquad
\{\barf,\bw\} = \frac{(n^2-1)\bb^2}{2n^2\bw^2(\bb-\bw)^2} .
\end{align}
The metric \eqref{schmet} after plugging in these expressions gives coordinates on the quotient manifold ${\cal \widetilde{M}}_n = {\cal M}_n / \mathbb{Z}_n$, since its boundary is just one copy of the $w$-plane. The full replica manifold ${\cal M}_n$ is simply vacuum AdS$_3$ in the $z$-coordinate \eqref{poinz}. This is the holographic dual of the Cardy-Calabrese construction \cite{Hung:2011nu} (see also \cite{Faulkner:2013yia, Colin-Ellerin:2021jev} for additional details).

\begin{figure}
\begin{center}
\begin{minipage}[b]{0.4\linewidth}
\flushleft
	\begin{overpic}[scale=0.35]{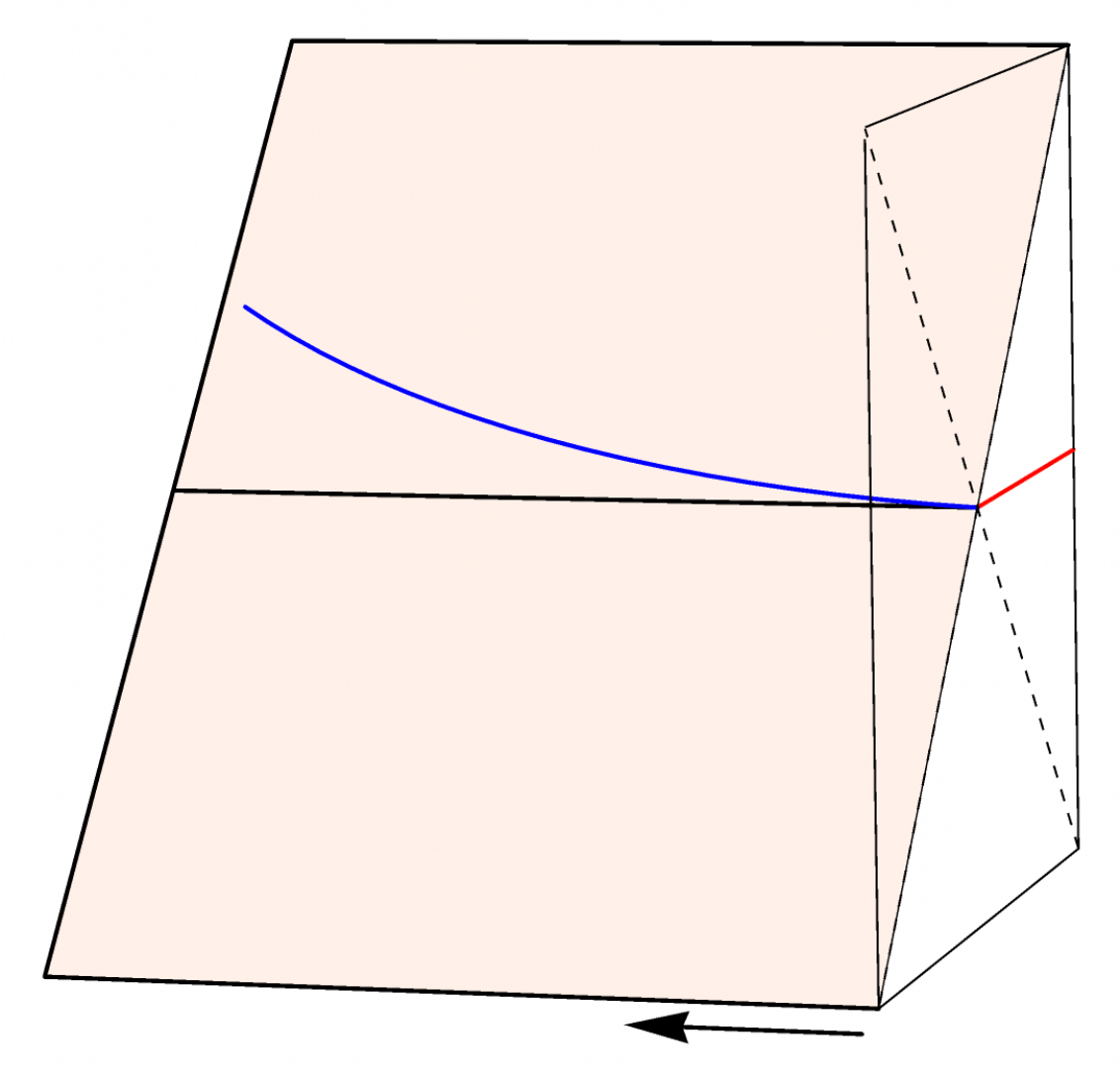}
\put(65,-2){\parbox{0.4\linewidth}{
		$\zeta$}}
\put(30,30){\parbox{0.4\linewidth}{
		$U = 0$}}
\put(52,60){\parbox{0.4\linewidth}{
		\textcolor{blue}{$V = e^x$}}}
\put(80,80){\parbox{0.4\linewidth}{
		$u$}}
\put(90,82){\parbox{0.4\linewidth}{
		$v$}}
\put(91,49){\parbox{0.4\linewidth}{
		\textcolor{red}{$A$}}}
	\end{overpic}
 \end{minipage}
\begin{minipage}[b]{0.5\linewidth}
\flushright
\begin{overpic}[scale=0.45]{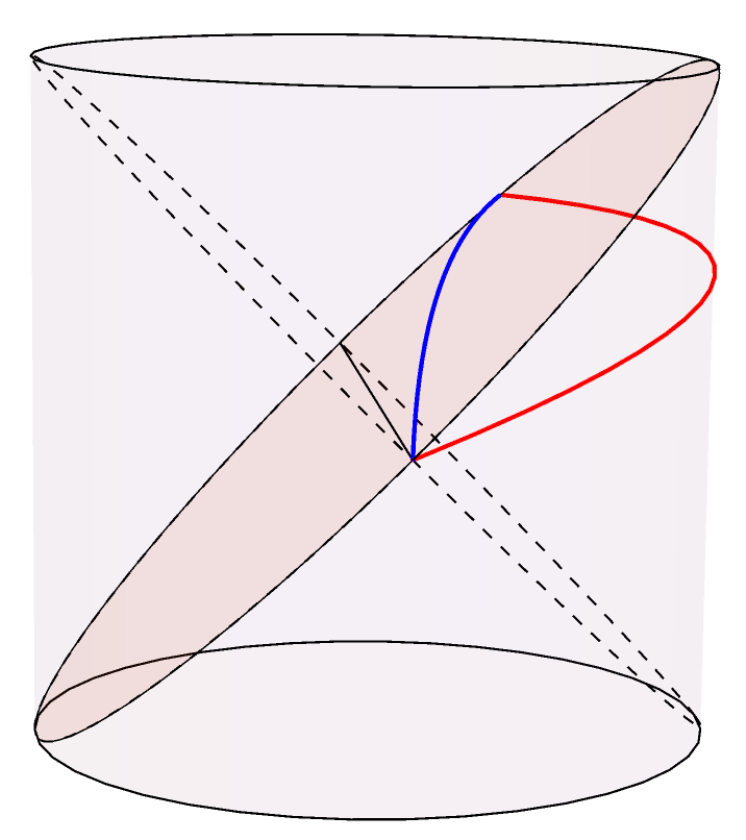}
\put(27,40){\parbox{0.4\linewidth}{
		$U = 0$}}
\put(85,65){\parbox{0.4\linewidth}{
		\textcolor{red}{$A$}}}
\put(42,78){\parbox{0.4\linewidth}{
		\textcolor{blue}{$V = e^x$}}}
\end{overpic}  
  \end{minipage}
\end{center}
\caption{ 
\small
Boundary region $A$ is a null-deformed half space in a two-dimensional CFT, with endpoints at $(0,0)$ and $(b,\infty)$ in null Minkowski coordinates $(u,v) = (-z,\bz)$. The bifurcation surface of the planar BTZ black hole is the black line at $U=V=0$. The entanglement surface (blue) for region $A$ lies on the null plane $U=0$, with a shift in the $V$-direction away from the bifurcation surface. This shift simultaneously determines the linearized replica manifold for region $A$,  the pole-skipping mode, and the shockwave metric. The right figure shows the same setup in global coordinates. 
\label{fig:btz}
}
\end{figure}
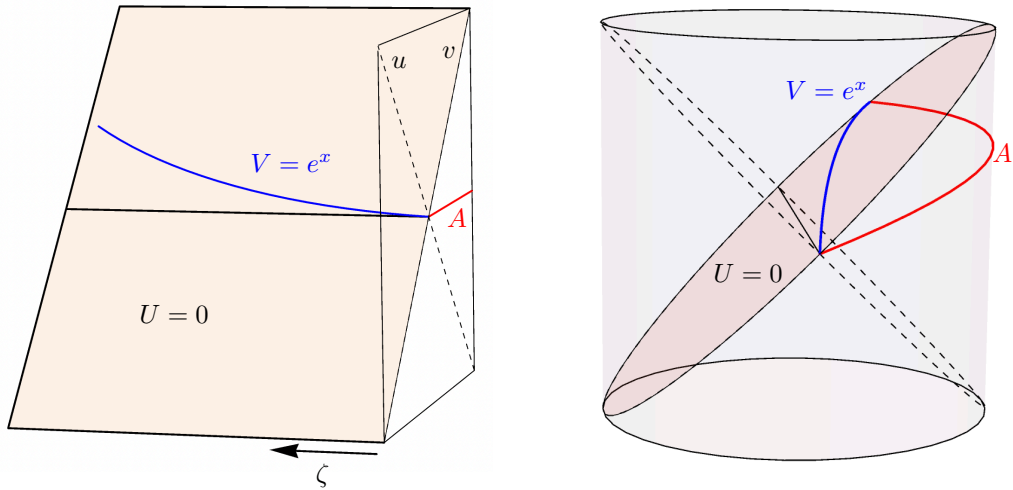

\subsubsection{Replica manifold of the null-deformed half-space}
This discussion was in Euclidean signature, but it immediately extends to Lorentzian signature by treating $(w,\bw)$ and $(b,\bb)$ as pairs of independent real numbers. Let us now set $\bb = \infty$ and $b \in \mathbb{R}$. The corresponding region $A$ is a null-deformed half-space. One twist operator is inserted at the origin and the other is on null infinity, at a location determined by $b$, such that $b \to \infty$ is the undeformed half-space. See figure \ref{fig:btz} for this region on the boundary and the corresponding entanglement wedge in the bulk, in Poincar\'e and global coordinates. 

We will denote the metric on the replica manifold for the region $A(b,\infty)$ by $ds_n^2(b)$. Taking $\bb \to \infty$ in \eqref{schmet} gives\footnote{The same result \eqref{quo} can be obtained without the need to take any limits by choosing 
\begin{align*}
f(w) = \left( \frac{w}{b-w}\right)^{1/n} , \quad \barf(\bw) = \bw^{1/n}. 
\end{align*}}
\begin{align}\label{quo}
\begin{split}
ds_n^2(b)
 &= \frac{dwd\bw+d\zeta^2}{\zeta^2}
  - \frac{(n^2-1)b^2}{4n^2w^2(b-w)^2} dw^2
   - \frac{(n^2-1)}{4n^2 \bw^2} d\bw^2
    + \frac{\zeta^2}{16} \frac{(n^2-1)^2 b^2}{n^4w^2 (b-w)^2 \bw^2} dw d\bw\\
    &= ds_n^2(\infty) - \frac{n^2-1}{4n^2} \frac{(2b-w)}{w(b-w)^2}dw^2
 +\zeta^2 \frac{(n^2-1)^2}{16n^4} \frac{(2b-w)}{w \bw^2 (b-w)^2 } dw d\bw
\end{split}
\end{align}
where $ds_n^2(\infty)$ is the replica manifold for an undeformed half-space,
\begin{align}\label{quoInf}
ds_n^2(\infty) = \frac{dwd\bw+d\zeta^2}{\zeta^2} - \frac{n^2-1}{4n^2 w^2} dw^2
 - \frac{n^2-1}{4n^2 \bw^2}d\bw^2 + \frac{\zeta^2}{16} \frac{(n^2-1)^2}{n^4 w^2 \bw^2}dw d\bw.
\end{align}
At leading order in $(n-1)$ things simplify as follows. With the twist operator at infinity:
\begin{align}\label{inflead}
ds_n^2(\infty) &=  \frac{dwd\bw+d\zeta^2}{\zeta^2} -(n-1) \frac{dw^2}{2w^2} - (n-1) \frac{d\bw^2}{2\bw^2} + \mathcal{O}((n-1)^2)
\end{align}
and for general $b$,
\begin{align}\label{genbleading}
\begin{split}
ds_n^2(b) &= 
 \frac{dwd\bw+d\zeta^2}{\zeta^2} - \frac{ (n-1) b^2}{2w^2(b-w)^2} dw^2 - (n-1) \frac{d\bw^2}{2\bw^2} + O((n-1)^2)\\
 &= ds^2(\infty)  - (n-1) \frac{(2b-w)}{2w(b-w)^2} dw^2 + O((n-1)^2).
 \end{split}
\end{align}

\subsubsection{Displacement mode}
The null displacement mode is by definition the change in the replica manifold under a linearized null deformation of the half-space $A(\infty,\infty)$, {\it i.e.}, the leading correction in $1/b$. Taking a derivative of \eqref{quo} and expanding at large $b$ we find 
\begin{align}
\p_b ds_n^2(b) &= \frac{n^2-1}{2n^2 b^2} \frac{dw^2}{w}-\zeta^2\fr{(n^2-1)^2 }{8n^4 b^2} \fr{dw d\bw}{w \bw^2} + \mathcal{O}(b^{-3}).
\end{align}
The displacement mode with one point at infinity is defined with the normalization \cite{Balakrishnan:2016ttg}
\begin{align}
\Delta \equiv b^2 \!\! \left. \frac{\p}{\p b}\right|_{b=\infty}.
\end{align}
Thus the displacement  mode for finite R\'enyi index $n$ is 
\begin{align}
\Delta ds_n^2 &= \frac{n^2-1}{2n^2} \frac{dw^2}{w} -\zeta^2\fr{(n^2-1)^2 }{8n^4 } \fr{dw d\bw}{w \bw^2}. 
\end{align}
Expanding to leading order in $(n-1)$, the displacement mode is\footnote{We use $\Delta$ for the displacement mode at finite $n$, so $\Delta ds^2_n = \delta ds^2_n + O((n-1)^2)$ where $\delta ds^2_n$ is the perturbation discussed in section \ref{sec:ps_equals_displacement}.}
\begin{align}\label{vacdisp}
\Delta ds_n^2 = (n-1) \frac{dw^2}{w} + O((n-1)^2). 
\end{align}
This agrees with the result of \cite{Balakrishnan:2016ttg} where the result in general dimensions was obtained by matching to the known stress tensor in the presence of a R\'{e}nyi defect.\footnote{Note that typically the displacement mode is defined as an infinitesmal local deformation of the region, but we consider a uniform deformation of the entire boundary of a ball-shaped region. In two dimensions, twist operators are points so these definitions coincide.}

The metric perturbation in \eqref{vacdisp} is a linearized solution to the Einstein equations that we obtained from the replica construction. It can also be viewed as a Rinder-space analogue of the pole-skipping mode after continuing to Lorentzian signature. This is a simple version of the basic relationship between the deformation mode and the pole-skipping mode that is the main point of this paper, though of course the nontrivial case is the black hole in higher dimensions.

\subsection{Replica vs. pole skipping for BTZ}
We will now translate this result into the language of the BTZ black hole, which is simply vacuum AdS$_3$ in different coordinates. This will connect it directly to the general discussion in section \ref{sec:ps_equals_displacement}.

\subsubsection{Branching around the bifurcation surface}
The planar BTZ black hole in Kruskal coordinates is
\begin{align}
ds^2 = \frac{-1}{(1+UV/4)^2} dUdV + \frac{(1-UV/4)^2}{(1+UV/4)^2} dx^2.
\end{align}
The AdS boundary is at $UV=-4$.\footnote{Note that we chose a different normalization of the Kruskal coordinates, compared to section \ref{sec:psshock}. }
This metric is related to vacuum AdS$_3$ \eqref{poinz} by the coordinate change
\begin{align}\label{ztobtz}
z =   \frac{-Ue^x}{1-UV/4}  \ , \qquad
\bz = \frac{Ve^x}{1-UV/4} \ , \qquad
\xi = \frac{1+UV/4}{1-UV/4}e^x. 
\end{align}
We start by describing the replica manifold when region $A$ is the entire right boundary, which is the region $A(\infty,\infty)$ in the notation used above. The corresponding Ryu-Takayanagi surface is the bifurcation surface $U=V=0$. The quotient manifold ${\cal \widetilde{M}}_n$ is therefore $ds_n^2(\infty)$ in \eqref{quoInf}. To map this to black hole coordinates, we compose \eqref{ztobtz} with
\begin{align}\label{bifchange}
z=w^{1/n} \frac{4n^2w\bw+(n^2-1)\zeta^2}{4n^2w\bw+(n-1)^2\zeta^2}, \quad
\bz=\bw^{1/n}\frac{4n^2w\bw+(n^2-1)\zeta^2}{4n^2w\bw+(n-1)^2\zeta^2}, \quad
\xi = \zeta\frac{4n (w\bw)^{\frac{n+1}{2n}}}{4n^2w\bw+(n-1)^2\zeta^2}.
\end{align}
This is \eqref{eq:roberts} with $f(w) = w^{1/n}$, $\barf(\bw) = \bw^{1/n}$, which places twist operators at $(w,\bw) = (0,0)$ and $(w,\bw) = (\infty,\infty)$. 
This shows that the replica manifold is identical to the original black hole,
\begin{align}\label{quoBtz}
    ds_n^2(\infty)
    &= 
    \frac{-1}{(1+UV/4)^2} dUdV + \frac{(1-UV/4)^2}{(1+UV/4)^2} dx^2. 
\end{align}
The only difference between the original black hole ($n=1$) and the replica manifold ($n \neq 1$) is in the coordinate change that takes $(U,V,x) \to (w,\bw,\zeta)$. This is simply because replicating around the bifurcation surface is the same as changing the temperature, and the metric of the planar black hole does not depend on temperature. Note that $(U,V,x)$ cover the full replica manifold ${\cal M}_n$, whereas $(w,\bw,\zeta)$ are coordinates on the quotient ${\cal \widetilde{M}}_n$.

To find the null-deformed replica manifold, we perform the same coordinate change (\eqref{bifchange} followed by \eqref{ztobtz}) on the metric $ds_n^2(b)$ in \eqref{genbleading}. The result, to first order in $(n-1)$, is
\begin{align}
    ds_n^2(b) &= \frac{-1}{(1+UV/4)^2} dUdV + \frac{(1-UV/4)^2}{(1+UV/4)^2} dx^2
    -(n-1) \frac{(2b-z)}{2z(b-z)^2} dz^2
     + \mo((n-1)^2)
\end{align}
where $z = \frac{-U e^x}{1-UV/4}$. Finally, to find the displacement mode, we keep only the leading term as $b \to \infty$:
\begin{align}\label{dispFullBtz}
\begin{split}
    ds_n^2(b) &= 
     \frac{-1}{(1+UV/4)^2} dUdV + \frac{(1-UV/4)^2}{(1+UV/4)^2} dx^2\\
     &\qquad 
    + (n-1) \frac{e^x}{b U (1-U V/4)^3}\Big( dU+U(1-UV/4)dx+\frac{1}{4}U^2 dV\Big)^2 \\
    &\qquad + \mo((n-1)^2, b^{-2} )
\end{split}
\end{align}
Near the horizon, $U \to 0$, the solution is
\begin{align}\label{nhbtz}
\begin{split}
    ds_n^2(b) &= 
     \frac{-1}{(1+UV/4)^2} dUdV + \frac{(1-UV/4)^2}{(1+UV/4)^2} dx^2
      \\
      & \hspace{30mm} + \frac{n-1}{b} F(x) \frac{dU^2}{U}  + \mathcal{O}((n-1)^2,b^{-2},U^0),
\end{split}
\end{align}
with $F(x) = e^x$,
which is the solution to the homogeneous shockwave equation in AdS$_3$ showing $v_B=1$ \cite{Ramirez:2020qer}. In \eqref{nhbtz} we recognize the perturbation as the pole-skipping mode near the horizon. The exact perturbation in \eqref{dispFullBtz} is the full pole-skipping mode that one obtains by the procedure described in section \ref{sec:psshock}, but in a different gauge. The following coordinate change brings the solution into radial gauge:
\begin{align}
\begin{split}
U &\to U - \frac{(n-1)F(x) U^2}{4b}\left( \frac{2-UV}{(1-UV/4)^2}+\log(1-UV/4)\right) \\
V &\to V - \frac{(n-1)F(x)}{b}\left( \frac{3-\frac{5}{2}UV+\frac{1}{32}U^3V^3+\frac{1}{256}U^4V^4}{2(1-UV/4)^2}+\log(1-UV/4)\right)\\
x &\to x - \frac{(n-1)F(x) U}{4b}\left( \frac{6-2UV+\frac{3}{8}U^2V^2+2(1-UV/4)^2\log(1-UV/4)}{(1-UV/4)^3}-1\right).
\end{split}
\end{align}
This puts the replica manifold into the form 
\begin{align}
\begin{split}
    ds_n^2(b) &=  \frac{-1}{(1+UV/4)^2} dUdV + \frac{(1-UV/4)^2}{(1+UV/4)^2} dx^2
     \\
     & \hspace{30mm} + \frac{n-1}{b} F(x) \left( \frac{dU^2}{U} + dUdx\right)
      +  \mathcal{O}((n-1)^2,b^{-2}),
\end{split}
\end{align}
from which we can extract the displacement mode
\begin{equation}
    \Delta ds_n^2 = -(n-1) F(x) \left( \frac{dU^2}{U} + dUdx\right) + \mathcal{O}((n-1)^2),
\end{equation}
which precisely takes the form of the pole-skipping mode. This equation illustrates our main result, in the simple case of three-dimensional gravity: The replica manifold branched around a shifted cut of the horizon, to leading order in $(n-1)$ and the deformation $b^{-1}$, is given by the background solution plus the pole-skipping mode. This case is a bit trivial because all solutions to the equations of motion in three-dimensional gravity are locally equivalent, but it provides an exactly solvable example of the general argument in section \ref{sec:ps_equals_displacement}.

\ \\
\ \\
\noindent\textbf{Acknowledgments}\\
\noindent We thank Mike Blake, Horacio Casini, Xi Dong, Tom Faulkner, Yikun Jiang, Hong Liu, Juan Maldacena, Mark Mezei, Mukund Rangamani, Douglas Stanford, Shreya Vardhan, Diandian Wang and Zhencheng Wang for helpful discussions. TH and WZC are supported by NSF grant PHY-2014071. WWW is supported by  DOE grant DE-SC0020397. We also acknowledge the Aspen Center for Physics where some of this work was completed supported by NSF grant PHY-2210452.

\bibliographystyle{ourbst}
\bibliography{butterfly.bib}
\end{document}